# Correlated Photon Pair Production by Spontaneous Parametric Down Conversion in Quasi-Phase-Matched AlGaAs superlattice Waveguides using a Continuous Wave Pump

**Peyman Sarrafi[1], Eric Zhu[1], Ksenia Dolgaleva[1], Barry M. Holmes[2], David C. Hutchings[2], Stewart Aitchison[1], Li Qian[1]**

[1]*Dept. of Electrical and Computer Engineering, Univ. of Toronto, 10 King's College Road, Toronto, Ontario, Canada M5S 3G4*

[2]*School of Engineering, University of Glasgow, Glasgow G12 8QQ, Scotland, U.K.*
peyman.sarrafi@utoronto.ca

**Abstract**

We report on the demonstration of correlated photon pair generation in quasi-phase-matched superlattice AlGaAs waveguides with a high coincidence-to-accidental ratio (CAR); with a continuous (CW) pump, the observed CAR (>100) is more than two order of magnitudes improvement over previously reported spontaneous down conversion (SPDC) schemes in AlGaAs waveguides with a pulsed pump.

---

Correlated photon pair sources are of primary importance in quantum communication [1] and quantum computing [2]. Many previous experimental demonstrations of these sources have been reported in crystal and waveguides[3,4], fibres[5,6], and silicon waveguides [7-9]. There is considerable interest in III-V semiconductors as a platform due to the prospect of potential monolithic integration with a diode laser pump thus obtaining correlated photon pair production from a single electrically powered chip.

There are two main schemes to generate correlated photon pairs in nonlinear waveguides: by using spontaneous four-wave-mixing (SFWM), a third-order nonlinear process relying on $\chi^{(3)}$ nonlinearity, or by utilizing spontaneous parametric down conversion (SPDC), also known as parametric fluorescence, based on $\chi^{(2)}$ nonlinearity.

The challenges with the SFWM scheme include the suppression of Raman scattering noise [10] and the suppression of pump light because the pump frequency and the signal/idler frequencies are normally all located within a limited spectral range. In contrast, SPDC-based correlated photon sources emit at around half the frequency of the pump. SPDC in the 1550 nm telecommunications band in AlGaAs-based waveguides offers the prospect of using well-developed fabrication technologies to incorporate laser pump source integration and to use the comparatively large second-order nonlinearity, which makes it a promising candidate for highly integrated photon pair sources.

Since AlGaAs lacks natural birefringence, in order to obtain efficient SPDC in the material, one has to devise a method for phase matching. This has been achieved using Bragg reflection waveguides [11] or vertical quasi-phase matching (QPM) by alternating AlGaAs layers with different aluminum concentration [12]. However, the structure used in [11] resulted in poor coincidence-to-accidental ratio (CAR) and the structure in [12] cannot be implemented in a planar integrated photonic circuit. In addition, both the demonstrations reported in [11] and [12] required a mode-locked picosecond pump source with high pulse energy.

In this letter, we report on a low noise, high CAR (>100), CW pumped correlated photon pair source based on quasi-phase-matched AlGaAs superlattice waveguides. This work is also the first demonstration of a correlated photon pair source that can be easily monolithically integrated with on-chip pump laser sources fabricated on similar superlattice substrate [13].

The correlated photon pair source is based on an AlGaAs waveguide which produces photon pairs via $\chi^{(2)}$ nonlinearity. A Type I QPM is achieved through quantum-well intermixing (QWI) [14], along the propagation direction. QWI modulates the energy band gap of the superlattice and, hence, the magnitude of $\chi^{(2)}$. This method of in-plane-phase matching offers relatively low scattering propagation loss. Furthermore, the phase-matching of the SPDC permits both the correlated photon pairs and pump to be in the fundamental guiding mode of the waveguide allowing for efficient coupling into fibres and straightforward integration with other photonic devices. This type of phase matching has been previously utilized for difference frequency generation and second harmonic generation (SHG) [15-17].

The device consists of a waveguide with a 0.6$\mu$m-thick core layer of 14:14 monolayer GaAs/Al$_{0.85}$Ga$_{0.15}$As superlattice, buffer layers of 300nm Al$_{0.56}$Ga$_{0.44}$As on both sides, and cladding layers of 800nm Al$_{0.60}$Ga$_{0.40}$As. There is an additional 1-$\mu$m-thick layer of Al$_{0.85}$Ga$_{0.15}$As underneath the lower cladding in order to avoid field leakage to GaAs substrate. The choice of Al concentration in the core layer of the AlGaAs waveguides results in low two photon absorption and good mode confinement for the 1550 nm band. Spatially periodic ion-implantation induced QWI in the superlattice was fabricated as follows. A layer of PMMA was initially spin-coated on top of AlGaAs substrate. Then the gratings with different periods were defined in the PMMA mask by an electron beam lithography. After that, a 2.3 $\mu$m-thick Au implantation mask was deposited in the PMMA gaps by electroplating with subsequent removal of PMMA. The ion implantation was carried out with 4.0 MeV $As^{2+}$ ions at a dosage of $2.0 \times 10^{13}$ $cm^{-2}$ to create point defects, followed by removal of the Au mask and rapid thermal annealing (RTA) at 775 °C for 60 s to provide QWI through the point defect diffusion.

Ridge waveguides of different widths with the height of 1 μm were fabricated by reactive ion etching, and the resulting sample was cleaved to a length of 3.5 mm. For this letter, a 3 μm-wide waveguide with a periodicity of 3.5μm was used. A schematic representation of the device is depicted in Fig. 1.

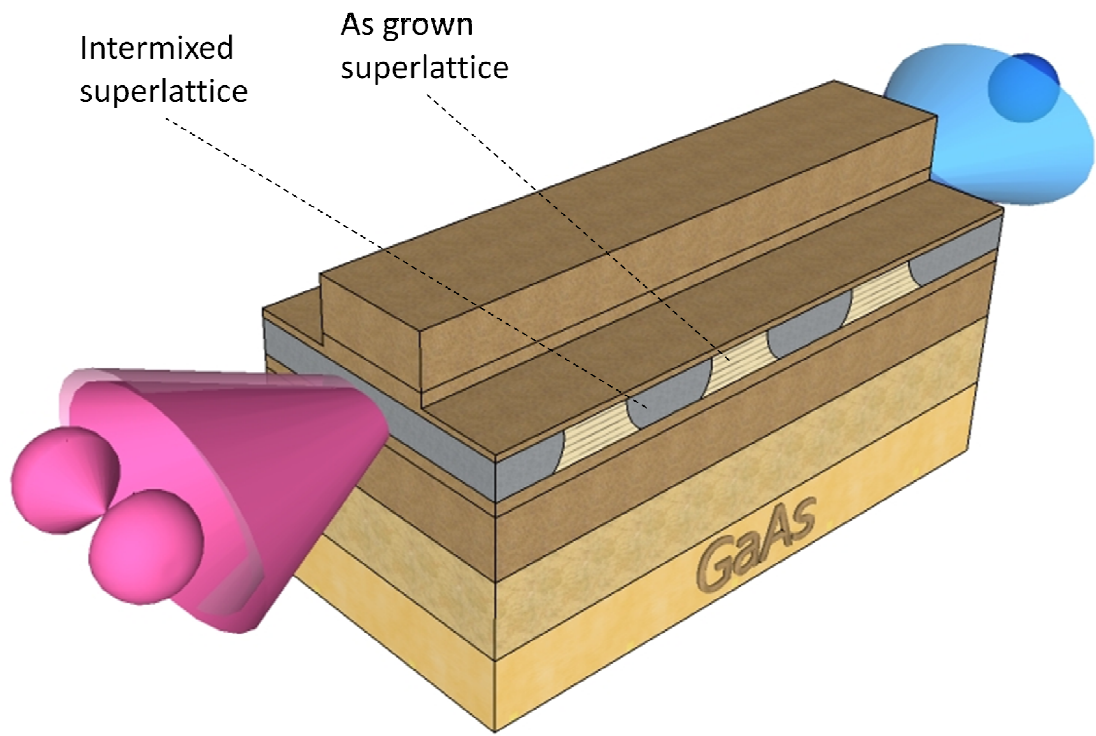

Fig. 1. A schematic of Quasi-Phase-Matched AlGaAs superlattice waveguides.

Correlated photon pair generation experiment and coincidence count were carried out using the experimental setup depicted in Fig. 2. A TM-polarized 772.8 nm CW source was used as a pump source for producing parametrically down converted correlated photon pairs, centered around double the pump wavelength 1545.6nm. This light was filtered by a polarizer (passing TE polarization only), a long-pass filter, and three cascaded fiber-based pump suppression filters. The down-converted photons were then deterministically separated into two spectral bands (1562–1578 nm and 1522–1538 nm) using two fiber-pigtailed bandpass filters (BPFs) with 16 nm bandwidths, centered around 1570 and 1530 nm. The optimal BPFs, were are not available, and the existing filters were not frequency-conjugate about the degenerate phase matching wavelength, and consequently correlated photons can only be detected across half the bandwidth of the BPFs (1562–1569.2 nm and 1522–1529.2 nm), reducing the achievable signal to noise ratio (CAR).

Two free-running id Quantique id220 single photon detectors (SPDs), with measured quantum efficiencies of 20% at 1550nm were used for coincidence measurements. The total estimated loss from the input objective to coupling-out fiber was 17 dB at 1550 nm. There is also a 5 dB loss associated with the coupling of the pump into waveguide. The pump suppression filters had a total loss of 4.9 dB. The insertion losses of the 1570 and 1530 nm BPFs are 0.67 and 0.97 dB, respectively. Finally, an FPGA-based time interval analyzer (TIA) allows coincidence detection to be measured as a function of the difference in arrival time between SPD signals.

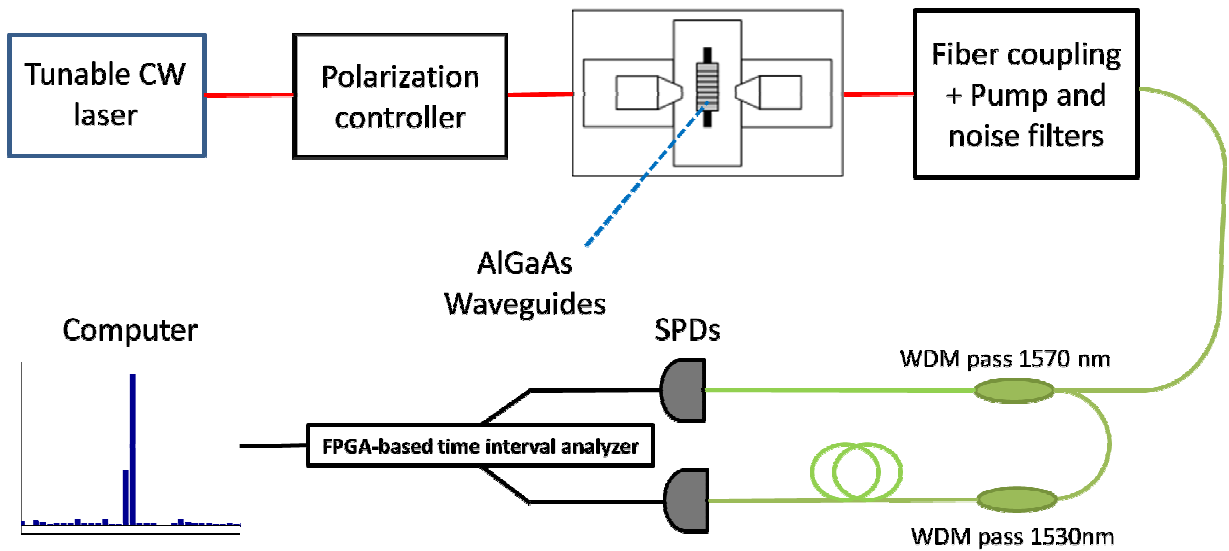

Fig. 2. Schematic of the setup for coincidence measurement in QPM AlGaAs superlattice waveguides.

A histogram of the raw coincidence measurements with a time-bin of 500 $ps$ is depicted in Fig. 3 (a). Here, the input CW pump power inside the sample is estimated to be ~8mW, and the integration time was 200 seconds. Coincidences versus the difference in arrival time is plotted and a peak at 26 ns is observed which corresponds to the relative optical path delay of the two arms of the BFPs, demonstrating true coincidence counts due to SPDC. The finite width of the peak is due to the electronic jitter of the detection system. The total electronic jitter is estimated to be 380 ps. A detailed study of electronic jitter has shown that each detector had a 200-250 ps jitter, and there was another 200 ps jitter due to the FPGA. As shown in Fig. 3(a), true coincidence is much higher than the accidental coincidence (including the contributions from the dark counts and all other noise photons), and a high CAR (>100:1) was measured. This is more than two order of magnitudes greater than the CAR previously reported using a spontaneous down conversion (SPDC) schemes in AlGaAs waveguides with a pulsed pump [11] while this source is introducing higher brightness.

In Fig. 3 (b), the CAR is plotted versus pump power (blue line). One can observe a positive slop for the low input powers. In order to interpret this result, the theoretical CAR formula for a pure source of correlated photons where there is no fluorescence or other type of noise is derived. This derivation is based on a number of survived photons-combination of a set of time-bins and calculating the associated probably of coincidence:

$$CAR = 0.45 \frac{\alpha n l^2}{(\alpha n l + d)^2 R},$$

where, $\alpha$ is the conversion efficiency, $n$ is the number of pump photons per second, $l$ is ratio of detected photons to generated photons ($10^{-3}$, corresponding to the overall loss coefficient), $d$ is detector's dark count (measured to be around 2000 counts per second) and $R$ is the resolution of the TIA (500 ps). The factor of 0.45 accounts for the reduction in effective bandwidth for coincidence measurements due to the non-optimal BPFs. The conversion efficiency $\alpha = 6 \times 10^{-11}$ pairs/pump photon for a 7.2nm effective bandwidth (corresponding a brightness of $1.8697 \times 10^{+6}$ down converted photon pairs per second in mentioned effective bandwidth and pump power of 8mW) has been extracted by fitting the two curves to have the maximum CAR at the same coupled pump power value. An excellent agreement between theory and measurement is obtained despite the fact no other normalization factor is used.

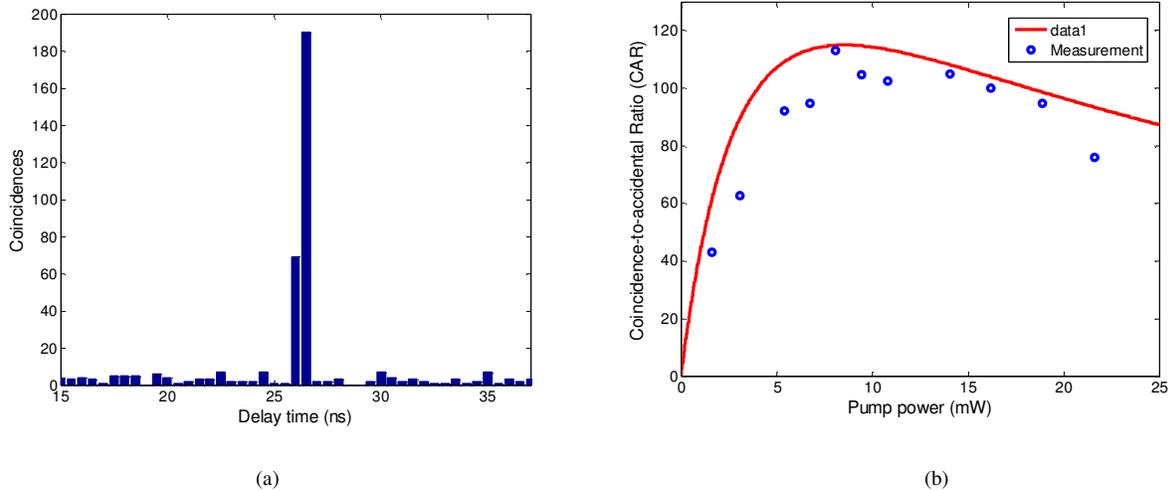

Fig. 3. (a) The typical time-bin histogram for coincidence measurements. (b) CAR as a function of coupled pump power.

To confirm that the generated photon pairs are due to type-I phase matching, the polarization of the down-converted photons were tested with a rotating polarizing beam splitter. When the output was TM-polarized, CAR drastically dropped to one, and there was no true coincidence observed. Additionally, the phase-matching bandwidth was measured by detuning the pump wavelength. The CAR measurement was performed as a function of the pump wavelength (for the pump power 6 mW). The results are shown in Fig. 4. The bandwidth and asymmetric dependence are in accord with the predicted tuning curves for DFG shown in reference [15]. Increasing the pump wavelength beyond the degeneracy point (corresponding to SHG) results in the phase-matching condition no longer being satisfied. Decreasing the pump wavelength from the degeneracy point by only around 1 nm, results in a tuning of the signal and idler wavelengths of ~100 nm and therefore they no longer coincide with the effective passband of the BPFs.

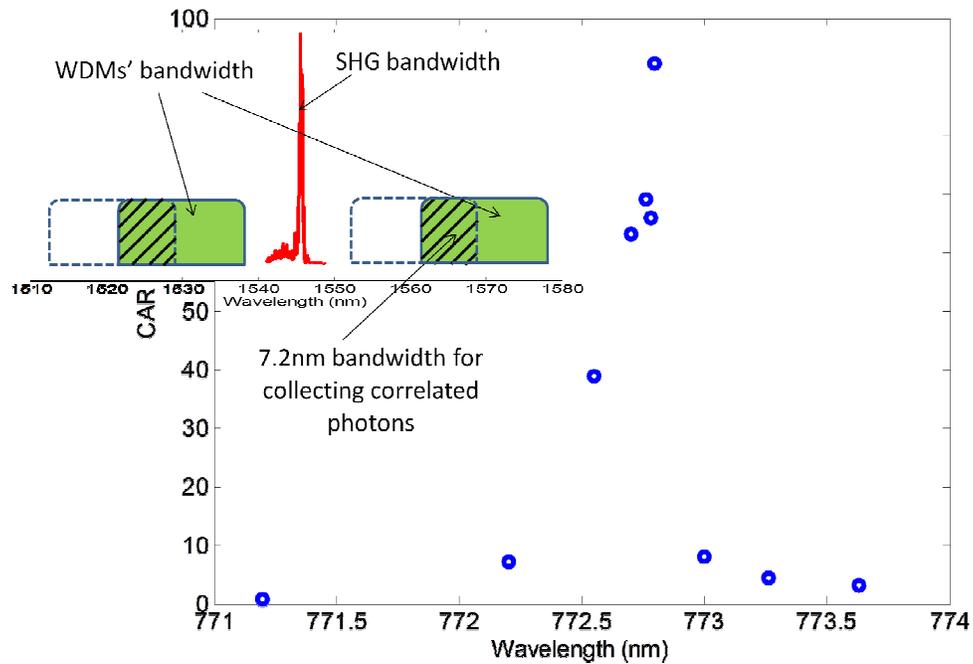

Fig. 4. Pump detuning. Figure inset: shows spectral location of collected pair photons (hatch pattern)

There is scope for improving the rate of production of the correlated photon pairs. Selecting BPFs that are frequency-conjugate about twice the pump wavelength would significantly improve the measured CAR. Using a 3dB coupler instead of 1570nm/1530nm BPFs result in about 8 times improvement in brightness of the SPDC source at the expense of lower CAR due to the greater increase in noise photon detection events. The number of total true coincidences per unit time can be significantly increased by improving the overall optical losses in the system and in particular end-fire coupling efficiencies.

In conclusion, the QWI quasi-phase-matching method in a superlattice AlGaAs waveguide has been used to implement a pure source of correlated pair photons with CW pumping. Our results demonstrate the feasibility of a CW-pumped, low-cost, planarly integrated, high-quality, turn-key photon pair source.

Acknowledgements

This work was supported by NSERC and authors want to acknowledge for supporting this research. We acknowledge ID Quantique for providing single photon detectors utilized in this work. The authors would like to thank the Ion Beam Centre, University of Surrey, Guildford, U.K., for carrying out the ion implantation. We also acknowledge the valuable contributions by the technical support staff of the James Watt Nanofabrication Centre at the University of Glasgow, Glasgow, U.K. This work was supported by the Engineering and Physical Sciences Research Council.

## References

[1] N. Gisin and R. Thew, "Quantum Communication", *Nat. Photonics* **1**, 165 (2007).
[2] M. A. Nielsen and I. L. Chuang, *Quantum Computation and Quantum Information* (Cambridge University Press, Cambridge, MA, 2000).
[3] M. Fiorentino, S. Spillane, R. Beausoleil, T. Roberts, P. Battle, and M. Munro, "Spontaneous parametric down-conversion in periodically poled KTP waveguides and bulk crystals," *Opt. Express* **15**, 7479-7488 (2007).
[4] Tanzilli, S.; De Riedmatten, H.; Tittel, H.; Zbinden, H.; Baldi, P.; De Micheli, M.; Ostrowsky, D.B.; Gisin, N.; , "Highly efficient photon-pair source using periodically poled lithium niobate waveguide," *Electronics Letters* , vol.37, no.1, pp.26-28, 4 Jan 2001
[5] Li, X.; Voss, P.L.; Sharping, J.E.; Kumar, P., "Optical-Fiber Source of Polarization-Entangled Photons in the 1550 nm Telecom Band" Phys. Rev. Lett., 94, 053601, 2005
[6] E.Y. Zhu, Z. Tang, et al., Physical Review Letters, **108**, 213902 (2012)
[7] Matsuda, Nobuyuki, Hanna Le Jeannic, Hiroshi Fukuda, Tai Tsuchizawa, William John Munro, Kaoru Shimizu, Koji Yamada, Yasuhiro Tokura, and Hiroki Takesue. "A monolithically integrated polarization entangled photon pair source on a silicon chip." *Nature Scientific Reports* 2 (2012).
[8] J. Sharping, K. Lee, M. Foster, A. Turner, B. Schmidt, M. Lipson, A. Gaeta, and P. Kumar, "Generation of correlated photons in nanoscale silicon waveguides," Opt. Express 14, 12388-12393 (2006).
[9] Takesue, H.; , "Entangled Photon Pair Generation Using Silicon Wire Waveguides," *Selected Topics in Quantum Electronics, IEEE Journal of* , vol.18, no.6, pp.1722-1732, Nov.-Dec. 2012
[10] C. Xiong, G. D. Marshall, A. Peruzzo, M. Lobino, A. S. Clark, D.-Y. Choi, S. J. Madden, C. M. Natarajan, M. G. Tanner, R. H. Hadfield, S. N. Dorenbos, T. Zijlstra, V. Zwiller, M. G. Thompson, J. G. Rarity, M. J. Steel, B. Luther-Davies, B. J. Eggleton, and J. L. O'Brien, "Generation of correlated photon pairs in a chalcogenide $As_2S_3$ waveguide", *Appl. Phys. Lett.*, 98, 051101 (2011)
[11] R. Horn, P. Abolghasem, B. J. Bijlani, D. Kang, A. S. Helmy, and G. Weihs, 'Monolithic Source of Photon Pairs', Phys. Rev. Lett., vol. 108, 153605 (2012)
[12] A. Orieux, X. Caillet, A. Lemaître, P. Filloux, I. Favero, G. Leo, and S. Ducci, "Efficient parametric generation of counterpropagating two-photon states," J. Opt. Soc. Am. B 28, 45-51 (2011).
[13] Younis, U.; Holmes, B.M.; Hutchings, D.C.; Roberts, J.S.; , "Towards Monolithic Integration of Nonlinear Optical Frequency Conversion," *Photonics Technology Letters, IEEE* , vol.22, no.18, pp.1358-1360, Sept.15, 2010
[14] J. H. Marsh, Quantum well intermixing, *Semicond. Sci. Technol.* **8**, 1136 (1993)
[15] S. J. Wagner, B. M. Holmes, U. Younis, I. Sigal, A. S. Helmy, J. S. Aitchison, and D. C. Hutchings, "Difference Frequency Generation by Quasi-Phase Matching in Periodically Intermixed Semiconductor Superlattice Waveguides," *IEEE Journal of Quantum Electronics*, vol. 47, pp. 834-840, 2011
[16] S. J. Wagner, B. M. Holmes, U. Younis, A. S. Helmy, J. S. Aitchison, and D. C. Hutchings, "Continuous wave second-harmonic generation using domain-disordered quasi-phase matching waveguides," *Appl. Phys. Lett.*, vol. 94, no. 15, pp. 151107-1–151107-3, Apr. 2009.
[17] D. Hutchings, S. Wagner, B. Holmes, U. Younis, A. Helmy, and J. Aitchison, "Type-II quasi phase matching in periodically intermixed semiconductor superlattice waveguides," Opt. Lett. 35, 1299-1301 (2010).